\documentstyle[12pt]{article}
\newlength{\extraspace}
\newlength{\extraspaces}
\newcommand{\be}{\begin{equation}}
\newcommand{\ee}{\end{equation}}
\newcommand{\br}{\begin{eqnarray}}
\newcommand{\er}{\end{eqnarray}}
\newcommand{\ba}{\begin{array}}
\newcommand{\ea}{\end{array}}
\newcommand{\bi}{\begin{itemize}}
\newcommand{\ei}{\end{itemize}}
\newcommand{\bn}{\begin{enumerate}}
\newcommand{\en}{\end{enumerate}}
\newcommand{\ul}{\underline}
\newcommand{\ol}{\overline}
\newcommand{\su}{\mbox{$SU(2)\times U(1)$}}
\begin{document}
\pagestyle{empty}
\begin{titlepage}
\begin{flushright}
UTPT-93-09
\\ hep-ph/9305337
\\ June 1993
\end{flushright}
\vspace{2.5cm}
\begin{center}
{\LARGE Metacolor}\\ \vspace{40pt}
{\large B. Holdom\footnote{holdom@utcc.utoronto.ca}}
\vspace{0.5cm}

{\it Department of Physics\\ University of Toronto\\
Toronto, Ontario\\Canada M5S 1A7}
\vskip 2.1cm
\rm
\vspace{25pt}
{\bf ABSTRACT}

\vspace{12pt}
\baselineskip=18pt
\begin{minipage}{5in}

A new mechanism is presented for the generation of quark and lepton masses,
based on a heavy fourth family and a new sector of massless fermions.  The
massless fermions have only discrete chiral symmetries and they are
confined by the metacolor force.  The resulting electroweak corrections may
be smaller than in technicolor theories.

\end{minipage}
\end{center}
\vfill
\end{titlepage}
\pagebreak
\baselineskip=18pt
\pagestyle{plain}
\setcounter{page}{1}

In technicolor theories, the dynamical technifermion mass has two functions.
One is to break electroweak symmetry and the other, in the presence of
suitable four fermion operators, is to generate mass for ordinary quark and
leptons. But with recent high precision data from $Z$ factories, these two
functions appear to be increasingly at odds. The way the data is
constraining the electroweak symmetry breaking sector is conveniently
described in terms of the $S$ and $T$ parameters~\cite{C}. The constraint
on $S$ limits the total number of massive technifermions (number of
techniflavors times number of technicolors) while the constraint on $T$
limits the amount of isospin breaking in the technifermion masses. In their
role of feeding down mass to the ordinary quark and leptons, technifermions
are being pushed up against these constraints. Models with minimal
extended-technicolor (ETC) interactions tend to have at least one complete
family of technifermions with standard $SU(3)\times SU(2)\times U(1)$
quantum numbers. And the large top mass tends to imply substantial isospin
symmetry breaking in technifermion masses~\cite{D}. One way to describe the
difficulty is that the same sector is responsible for both electroweak
symmetry breaking and fermion mass generation.

This naturally leads to the question of whether it is possible to separate
these two functions and associate each with its own sector. That is, is it
possible to have an electroweak breaking sector with a reasonably small
number of new heavy fermions and good weak isospin symmetry? This may be
possible if there was another sector which did not contribute to
electroweak breaking and yet was required for the generation of quark and
lepton masses. The isospin symmetry breaking present in this sector could
then feed into the quark and lepton masses without feeding into electroweak
corrections.

We shall attempt to realize this possibility in an economical fashion by
introducing a fourth family. Consider a theory with the following gauge
symmetries above the scale of electroweak symmetry breaking, where the new
gauge symmetry is hypercolor. \be{G}_{H}\times SU(3)\times SU(2)\times
U(1)\label{d}\ee Each fermion in the third and fourth families are within
hypercolor multiplets.  The two light families are hypercolor singlets, and
we shall return to them later. It will be important for what follows that
these gauge symmetries are the only continuous symmetries present.

The hypercolor coupling is assumed to be of order unity at $\sim 1$ TeV.
But we also consider the existence of nonrenormalizable interactions
generated by strong interaction physics at higher scales.  At the very
least these interactions must respect the gauge symmetries in (\ref{d}).
Because these scales of new physics may not be much higher than 1 TeV,
and/or because of nontrivial scaling effects which can occur in theories
with small $\beta$-function, it may be that some of these nonrenormalizable
interactions can not be treated as small perturbations to hypercolor
dynamics.  They may play an important role in the symmetry breaking
occurring at $\sim 1$ TeV.  We will refer to this subset of the full set of
nonrenormalizable interactions as the dominant nonrenormalizable (DNR)
interactions.

We will describe as much of the basic picture as possible before specifying
${G}_{H}$ and other details of the model.  First suppose that hypercolor
along with the DNR interactions produces electroweak symmetry breaking, but
in a channel which also breaks hypercolor.  We take this channel to
correspond to a mass for all members of the fourth family.  The unbroken
subgroup of hypercolor will be metacolor.  The former hypercolor multiplets
will now contain metacolor singlets, and these singlet fermions comprise
all of the third as well as the fourth families. Metacolor will not be
responsible for the breaking of electroweak symmetry. Instead it is the
fourth family which breaks electroweak symmetry, and we thus assume that
the fourth family condensates respect isospin symmetry to good
approximation.

Consider the effective theory after integrating out the fourth family and
massive hypercolor gauge bosons.  \su\ symmetry may be realized in the
usual manner by introducing the Goldstone fields via the standard matrix
$U(x)$, transforming as $(2,2)$ under $SU(2{)}_{L}\times SU(2{)}_{R}$.  The
coefficients in the resulting chiral Lagrangian can be expected to be
smaller than in a one family technicolor model, since in the latter case
the coefficients contain a factor of the number of technicolors.  In
particular the contribution to the parameter $S$ from $\sim 1$ TeV physics
will be smaller.  Pseudo-Goldstone bosons exist similar to the technipions
in one family technicolor, and their contribution to $S$ will be similar.

The {\em central assumption} in our picture is the following. \bi \item
Certain discrete symmetries of the gauge and DNR interactions remain
unbroken, and are such as to prevent the metafermions from receiving mass
of any kind.\ei

The model will illustrate how such discrete chiral symmetries can arise.

The effective theory below the hypercolor breaking scale thus includes
massless metafermions.  But we may assume that metacolor confines at a
scale somewhat below the hypercolor breaking scale and well above the $Z$
mass.  There are then no bound states of metafermions much lighter than
this confinement scale. The reason is simple: there are no continuous
chiral symmetries in the theory besides \su.  Note that at scales close to
the hypercolor breaking scale, hypercolor interactions badly break other
potential global chiral symmetries involving metafermions.  Thus there is
no symmetry reason for the existence of light states other than the
Goldstone bosons absorbed by the $W$ and $Z$.

It is clear at this stage that since metafermions are massless, the
contributions to the parameters $S$ and $T$ from the metacolor sector will
bear little resemblance to the usual contributions in technicolor
theories.  For example in QCD it is known that a massive quark loop does
well at saturating the observed parameters in the low energy chiral
Lagrangian~\cite{F}, especially if the momentum dependence of the quark
mass is included~\cite{E}.  The contributions to the low energy theory from
a sector of confined {\em massless} fermions could then look quite
different.

The generation of quark and lepton masses will rely on four-metafermion
condensates.  We will show that \su\ conserving metafermion condensates
combined with \su\ violating fourth family condensates are all that is
required to generate quark and lepton masses.

Before proceeding further we must describe the model in more
detail.\footnote{Certain aspects of this model may be found in previous
work~\cite{A}.} The two light families will continue to be ignored for the
moment.  We take the hypercolor group to be $SU(3{)}_{H}$ with the
following multiplets. \be\ba{ccc} &\hspace{.2in}~{\cal Q}_{V}~{\cal
Q}_{A}~{\cal Q}_{F}\\ {\bf \Psi }_{L}&(3,~+,~+,~+)\\ {\bf \Psi
}_{R}&(3,~+,~-,~+)\\ {\bf \ul \Psi }_{L}&(\ol{3},~-,~-,~+)\\ {\bf \ul \Psi
}_{R}&(\ol{3},~-,~+,~+)\label{j}\ea\ee $\bf \Psi $ and $\bf \ul \Psi $
represent two complete hypercolored families containing \boldmath$(U ,D ,N
,E)$\unboldmath\ and \boldmath$(\ul U ,\ul D ,\ul N ,\ul E )$\unboldmath\
respectively.  All these fermions have standard $SU(3)\times SU(2)\times
U(1)$ quantum numbers, and thus the only difference between the $\bf \Psi$
and $\bf \ul \Psi$ fermions is that they transform differently under
$SU(3{)}_{H}$. We have also defined three diagonal generators, ${\cal
Q}_{V}$, ${\cal Q}_{A}$, and ${\cal Q}_{F}$, which commute with all gauge
generators.  We will describe their connection with the discrete symmetries
below.

Hypercolor $SU(3{)}_{H}$ breaks to metacolor $SU(2{)}_{M}$.  The
decomposition $3\Rightarrow 2+1$ and $\ol 3\Rightarrow \ol 2+\ol 1$
corresponds to $\bf \Psi \rm \Rightarrow \Psi +\psi $ and $\bf \ul \Psi \rm
\Rightarrow {\ul \Psi} +{\ul \psi}$ respectively. $\Psi $ and $\ul\Psi $
represent two complete metacolored families containing $(U,D,N,E)$ and
$(\ul U,\ul D,\ul N,\ul E)$.  $\psi $ and $\ul\psi $ containing $(t,b,\nu
,\tau )$ and $(\ul t,\ul b,\ul \nu ,\ul \tau )$ will describe the third and
fourth families.  The actual mass eigenstates for the third and fourth
families are determined by the symmetry breaking.  The condensate which
breaks hypercolor and generates the fourth family mass is taken to be
\be\left\langle{{\ol{\bf\ul \Psi }}_{L}^{A}{\bf\Psi }_{R}^{B}}+{\rm
h.c.}\right\rangle\propto {\delta }^{A3}{\delta }^{B3}.\label{e}\ee $A$,
$B$ are $SU(3{)}_{H}$ indices.  This condensate lies in the $\bf 6$ of
$SU(3{)}_{H}$ and it transforms as $(2,2)$ under $SU(2{)}_{L}\times
SU(2{)}_{R}$.  We consider below possible DNR interactions which favor this
condensate.

Thus far we have considered the chiral Lagrangian obtained from integrating
out the fourth family.  The coefficients of the various terms may be
estimated from a naive one-loop diagram~\cite{F,E}, keeping in mind that
there are also hypercolor corrections.  We now want to consider possible
new contributions after integrating out the metacolor sector.  There is
very little we can say about the effects of the possible dynamical \su\
breaking in the metacolor sector.  But we can look at the effects of the
explicit \su\ breaking in the metacolor sector fed down from the fourth
family. This will show up as couplings between the $U(x)$ field and the
metafermions.  First consider the terms ${\ol{\Psi }}_{L}U{\Psi }_{R}$,
${\ol{\ul \Psi }}_{L}U{\Psi }_{R}$, ${\ol{\Psi }}_{L}U{\ul \Psi }_{R}$, and
${\ol{\ul \Psi }}_{L}U{\ul \Psi }_{R}$. It is easy to see from the gauge
structure and (\ref{e}) that these terms are not generated from hypercolor
gauge boson exchange.  This observation will be rephrased below in terms of
discrete symmetries, which if unbroken, prevent any kind of metafermion
mass term.  Had the metafermions received mass then they would have made
the usual contribution to the term ${\rm Tr}(U{W}_{\mu \nu }U^{\dagger
}{B}^{\mu \nu })$, whose coefficient is proportional to $S$~\cite{C}.

To find a metafermion contribution to $S$ we consider the following terms,
which are generated via a one-loop diagram involving a massive hypercolor
gauge boson and a massive fourth family fermion. \be {\ol{\ul{\Psi
}}}_{L}U({D}\!\!\!\!{/}\;{U}^{\dagger }){\ul{\Psi }}_{L}+{\ol{\Psi
}}_{R}{U}^{\dagger }({D}\!\!\!\!{/}\;U){\Psi }_{R}+{\rm h.c.}\label{m} \ee
$D_\mu$ is the usual weak covariant derivative.  These terms have a
coefficient which is of order ${\alpha }_{H}/2\pi$ where ${\alpha }_{H}$ is
the hypercolor coupling.  Because these terms induce, for example, a
coupling of the $W$ gauge boson to right-handed metafermions, a metafermion
loop can now generate a contribution to $S$.  But the origin of this
contribution, like (\ref{m}), involves the fourth family fermion mass and
it is proportional to ${\alpha }_{H}$.  This along with effects higher
order in ${\alpha }_{H}$ are really nothing but hypercolor corrections to
the fourth family contribution.  Lastly we note that ${\alpha }_{H}$ may be
of order unity or less, since DNR interactions are also playing a role in
the symmetry breaking at $\sim 1$ TeV.

Now we consider the origin of the third family masses.  Tree level diagrams
involving massive fourth family fermions and massive hypercolor gauge bosons
generate the following terms (with $SU(2{)}_{L}\times SU(2{)}_{R}$ indices
shown). \be{\ol{\ul \Psi }}_{L}^{{i}_{1}}{\ul \Psi }_{{i}_{2}R}{\ol{\Psi
}}_{L}^{{i}_{3}}{\Psi }_{{i}_{4}R}{\ol{\ul \psi }}_{R}^{{i}_{2}}{\psi
}_{{i}_{3}L}{U^{\rm T}}{}^{{i}_{4}}{}_{{i}_{1}}+{\rm h.c.}\label{a}\ee The
relevant \su\ {\em invariant} metafermion condensate is then
\be\left\langle{{\ol{\ul \Psi }}_{L}^{{i}_{1}}{\ul \Psi
}_{{i}_{2}R}{\ol{\Psi }}_{L}^{{i}_{3}}{\Psi
}_{{i}_{4}R}}\right\rangle=A{\varepsilon }^{{i}_{1}{i}_{3}}{\varepsilon
}_{{i}_{2}{i}_{4}}+B{\varepsilon }^{{i}_{1}{i}_{3}}{\left[{{\tau
}_{3}\varepsilon }\right]}_{{i}_{2}{i}_{4}}. \label{b}\ee   $\varepsilon$
is the antisymmetric $2\times 2$ unit matrix. The second term accounts for
the effects of isospin symmetry breaking occurring in the metacolor sector.
Combining (\ref{a}) and (\ref{b}) gives terms proportional to \be A{\ol{\ul
\psi }}_{R}\tilde{U} {\psi }_{L}+B{\ol{\ul \psi }}_{R}{\tau }_{3}\tilde{U}
{\psi }_{L}+{\rm h.c.}\label{c}\ee where $\tilde{U}=\varepsilon {U}^{\rm
T}\varepsilon $. These terms describe third family masses with isospin
splitting.\footnote{The operator ${\ol{\ul{\bf \Psi }}}_{R}{\bf \Psi
}_{L}{\ol{\bf \Psi }}_{R}{\ul{\bf \Psi }}_{L}+{\rm h.c.}$ is not generated
explicitly by gauge interactions, but it is allowed by the gauge symmetries
in (\ref{d}). Such an operator could be present among the nonrenormalizable
interactions, and it also contributes to the third family masses in the
presence of the condensate (\ref{e}).}  This isospin breaking can feed into
electroweak corrections, but this is just the usual heavy top quark
contribution to $\Delta \rho$.

Isospin symmetry breaking arises from the absence of $SU(2{)}_{R}$ symmetry
in the DNR interactions. The condensate, (\ref{b}), reflects this
$SU(2{)}_{R}$ symmetry breaking and feeds it into the quark and lepton
masses.  The point is that this condensate is different from the
condensates responsible for electroweak symmetry breaking, unlike the
situation in technicolor theories.  There is no longer a direct link
between the top-bottom mass splitting and the isospin splitting of heavier
(techni)fermions, and thus producing a large top mass in this picture
should be easier.  We also note that in technicolor theories there is
another possible contribution to $T$ coming directly from a $SU(2{)}_{R}$
violating ETC interaction, producing a direct contribution to the term
$[{\rm Tr}({\tau }_{3}{U}^{\dagger }{D}_{\mu }U){]}^{2}$.  This type of
contribution can still arise from our metacolor sector due to the terms in
(\ref{m}).

We now turn to the generators ${\cal Q}_{V}$, ${\cal Q}_{A}$, and ${\cal
Q}_{F}$ and consider the discrete symmetries,
\be\exp\left({{\frac{1}{3}}\pi in{\cal Q}_{V}}\right)\label{g}\ee \be
\exp\left({{\frac{1}{2}}\pi in\left[{\cal Q}_{A}\pm {\cal
Q}_{F}\right]}\right),\label{h}\ee where $n$ is an integer.  All
four-hyperfermion operators allowed by the gauge symmetries in (\ref{d})
respect these discrete symmetries, and thus it is fair to assume that these
are symmetries of the DNR interactions.  We note in passing that the
allowed four-hyperfermion operators actually leave unbroken a continuous
${\cal Q}_{V}$ symmetry.  This may be broken by the following
six-hyperfermion operators.\footnote{To see that these operators can be
\su\ invariant, note that \su\ invariants can be made from
\boldmath${\ol{Q}}_{L}{U}_{R}{L}_{L}$\unboldmath\ or
\boldmath${\ol{D}}_{R}{Q}_{L}{L}_{L}$\unboldmath} \be \left({{\varepsilon
}_{ABC}{\varepsilon }_{DEF}{\ol{\ul{\bf \Psi}}}^{A}{\bf \Psi }^{B}{\bf \Psi
}^{C}{\ol{\ul{\bf \Psi}}}^{D}{\bf \Psi }^{E}{\bf \Psi }^{F}+{\rm
h.c.}}\right)~~,~~\left(\bf \Psi \rm \leftrightarrow \ul{\bf \Psi
}\right)\label{k}\ee  $A$, $B$, $C$ are $SU(3{)}_{H}$ indices.  The
discrete symmetries of these operators are exactly those in (\ref{g}) and
(\ref{h}).

The symmetry breaking represented by the fourth family condensate (\ref{e})
has the following effect.  The discrete symmetry in (\ref{g}) is broken,
but the combination of ${\cal Q}_{V}$ and the ${\lambda }_{8}$ generator of
$SU(3{)}_{H}$ yields the unbroken discrete symmetry
\be\exp\left({{\frac{1}{2}}\pi in{\hat{\cal Q}}_{V}}\right).\label{i}\ee
${\hat{\cal Q}}_{V}$ has the same quantum numbers as ${\cal Q}_{V}$ for the
metafermions, but ${\hat{\cal Q}}_{V}$ has zero quantum numbers for the
third and fourth families. The discrete symmetry in (\ref{h}) remains
unbroken. The point of all this is to be able to say that the unbroken
discrete symmetries in (\ref{h}) and (\ref{i}) are sufficient to prevent
any mass term, either Dirac or Majorana, for any of the metafermions.

It is clear that the DNR interactions must play an important role in the
dynamics which ends up preserving these discrete symmetries.  We give here
an example of the possible origin of some of the relevant DNR operators.
Assume that ${\cal Q}_{A}$ is the generator of a strong gauged $U(1)_A$
which is broken at some scale above a TeV.  Such an interaction will
produce DNR operators of the form ${\ol{\bf \Psi }}_{L}{\bf \Psi
}_{R}{\ol{\bf \Psi }}_{R}{\bf \Psi }_{L}$ and ${\ol{\ul{\bf \Psi
}}}_{L}{\ul{\bf \Psi }}_{R}{\ol{\ul{\bf \Psi }}}_{R}{\ul{\bf \Psi }}_{L}$
with signs which resist the formation of the hypercolor singlet condensates
$\langle {\ol{\bf \Psi }}_{L}{\bf \Psi }_{R}\rangle $ and $\langle
{\ol{\ul{\bf \Psi }}}_{L}{\ul{\bf \Psi }}_{R}\rangle $.  The $U(1)_A$ gauge
boson exchange also produces the operator ${\ol{\ul{\bf \Psi }}}_{L}{\bf
\Psi }_{R}{\ol{\bf \Psi }}_{R}{\ul{\bf \Psi }}_{L}$ with a sign which
favors the hypercolor violating condensate in (\ref{e}).  Such DNR
interactions can therefore make the formation of this condensate a
plausible possibility.  If $U(1)_A$ is broken at a fairly high scale then
we must assume that some or all of these operators are sufficiently
enhanced by anomalous scaling effects.

These $U(1)_A$ induced interactions also resist the formation of the
metafermion condensates $\langle {\ol{\Psi }}_{L}{\Psi }_{R}\rangle $ and
$\langle {\ol{\ul{\Psi }}}_{L}{\ul{\Psi }}_{R}\rangle $.  But they do not
resist the condensates $\langle {\ol{\Psi }}_{L}{\ul{\Psi }}_{R}\rangle $
and $\langle {\ol{\ul{\Psi }}}_{L}{\Psi }_{R}\rangle $ which are also
metacolor singlets.  On the other hand the broken $SU(3{)}_{H}$
interactions, and in particular the massive ${\lambda }_{15}$ gauge boson,
generates the effective operators ${\ol{\Psi }}_{L}{\ul{\Psi
}}_{R}{\ol{\ul{\Psi }}}_{R}{\Psi }_{L}$ and ${\ol{\ul{\Psi }}}_{L}{\Psi
}_{R}{\ol{\Psi }}_{R}{\ul{\Psi }}_{L}$ with signs which resist these latter
condensates.  We thus see that the metacolor plus DNR interactions may bear
little resemblence to QCD-like dynamics, and our central assumption that
certain discrete symmetries remain unbroken becomes a dynamical
possibility.\footnote{In the last two paragraphs we have ignored the
possibility of color and/or charge violating bilinear condensates, all of
which break the discrete symmetries.  These condensates are of course
resisted by the color and electromagnetic forces, but they can also be
resisted by DNR operators of the form ${\ol{\bf \Psi }}_{L}^m{\ul{\bf \Psi
}}_{R}^{cn}{\ol{{\ul{\bf \Psi }}^{c}}}_{Rn}{\bf \Psi
}_{Lm}+(L\leftrightarrow R)$ where $m$ and $n$ denote either a color or a
lepton ``flavor".  Such operators, with the right sign, can have their
origin in a Pati-Salam $SU(4)$ interaction.}

We now briefly sketch how the two light families may fit into this
picture.  We will denote all members of the light families by $\chi $.
Operators of the standard ETC form $\ol{\bf \Psi }\ul{\bf \Psi}\ol{\chi
}\chi $, which could have fed mass down from the fourth family to the light
fermions, are not $SU(3{)}_{H}$ invariant.  They are therefore not among
the nonrenormalizable interactions.  They are also not generated by
hypercolor dynamics since the light families are $SU(3{)}_{H}$ singlets.
This means that operators involving both metafermions and fourth family
fermions are required to generate light fermion masses.  This is what is
desired, since the isospin breaking in the metacolor sector is forced to
feed into the light quark masses.

The following operators are allowed by the gauge symmetries in
(\ref{d}).\footnote{These operators appear to break the discrete symmetries
in (\ref{g}) and (\ref{h}), but new discrete symmetries may be defined
which also act on light families.  We do not elaborate here.}
\be\renewcommand{\arraystretch}{.4}\ba{c} {\varepsilon }_{ABC}{\ol{\bf \Psi
}}_{\!\!\!\ba{c} \scriptstyle{L}\\ \scriptstyle{R}\ea}^{A}{\ul{\bf \Psi}
}_{\!\!\!\!\ba{c}\scriptstyle{R}\\ \scriptstyle{L}\ea}^{B}{\ol{\bf \Psi
}}_{R}^{C}{\chi }_{L}+{\rm h.c.}\\[20pt]{\varepsilon }_{ABC}{\ol{\bf \Psi
}}_{\!\!\!\ba{c}\scriptstyle{L}\\ \scriptstyle{R}\ea}^{A}{\ul{\bf \Psi}
}_{\!\!\!\!\ba{c}\scriptstyle{R}\\ \scriptstyle{L}\ea}^{B}{\ol{\chi
}}_{R}{\ul{\bf \Psi} }_{L}^{C}+{\rm h.c.}\ea\label{f}\ee The way the weak
indices are contracted depends on how the $L$ and $R$ subscripts are
attached.  Because of the symmetry breaking $SU(3{)}_{H}\Rightarrow
SU(2{)}_{M}$ we may replace these operators by the following.
\be\renewcommand{\arraystretch}{.4}\ba{c} {\ol{\Psi
}}_{\!\!\!\ba{c}\scriptstyle{L}\\ \scriptstyle{R}\ea}{\ul \Psi
}_{\!\!\!\!\ba{c}\scriptstyle{R}\\ \scriptstyle{L}\ea}{\ol{\psi }}_{R}{\chi
}_{L}+{\rm h.c.}\\[15pt]{\ol{\Psi }}_{\!\!\!\ba{c}\scriptstyle{L}\\
\scriptstyle{R}\ea}{\ul \Psi }_{\!\!\!\!\ba{c}\scriptstyle{R}\\
\scriptstyle{L}\ea}{\ol{\chi }}_{R}{\ul \psi }_{L}+{\rm h.c.}\ea\ee

When the fourth family is integrated out we obtain \be \ol{\Psi
}\ul{\Psi}\ol{\Psi }\ul{\Psi} {\ol{\chi }}_{R}{\chi }_{L}U^{\rm T}+{\rm
h.c.}\ee where there is again various ways to attach the $L$ and $R$
subscripts. There are appropriate four-metafermion condensates which are
\su\ invariant and which respect the discrete symmetries. The following
mass terms for the light quarks are induced. \be C{\ol{\chi }}_{R}\tilde{U}
{\chi }_{L}+D{\ol{\chi }}_{R}{\tau }_{3}\tilde{U} {\chi }_{L}+{\rm
h.c.}\ee  In terms of the various scales involved, these masses are of
order ${\Lambda }_{M}^{6}/{\Lambda }_{H}{\Lambda }_{NR}^{4}$ where
${\Lambda}_{NR}$ is the scale which characterizes the operators in
(\ref{f}).  To get reasonable light quark masses ${\Lambda}_{NR}$ cannot be
too high and/or there are additional powers of ratios of scales due to
anomalous scaling.  Note that it is the hypercolor interaction which would
be relevant for anomalous scaling, and that the operators in (\ref{f}) have
three hyperfermions.  Also, with the fermion content we have been
discussing the one-loop hypercolor $\beta$-function is in fact small.

Of special interest in our picture are the lepton masses. First we note
that the right-handed neutrino mass terms, ${\nu }_{eR}{\nu }_{eR}$, ${\nu
}_{\mu R}{\nu }_{\mu R}$, ${\nu }_{eR}{\nu }_{\mu R}$, and the right-handed
hyperneutrino mass terms,
${\mbox{\boldmath$N$}}_{R}{\ul{\mbox{\boldmath$N$}}}_{R}$, are allowed by
the gauge symmetries in (\ref{d}).  Let us suppose that all these masses
are much larger 1 TeV, generated at some high scale of ETC symmetry
breaking.  The large hyperneutrino mass removes the right-handed neutrinos
of the third and fourth families, and the right-handed metaneutrinos, from
the $\sim 1$ TeV effective theory.  The first question is the origin of the
(third family) $\tau$ mass, since the operators in (\ref{a}) can no longer
contain right-handed neutrinos or metaneutrinos.  But the following
operator is included in (\ref{a}) and it gives $\tau$ a mass in the
presence of the appropriate \su\ invariant metafermion condensate.\be
{\ol{\ul U}}_{L}{\ul E}_{R}{\ol{E}}_{L}{U}_{R}{\ol{\ul \tau }}_{R}{\tau
}_{L}{U}{}_{1}{}^{1}+{\rm h.c.}\ee  This operator is generated by a tree
level diagram involving a massive fourth family charge $2/3$ quark and
massive hypercolor gauge bosons.

The fourth family condensates cannot involve fields which are not present,
and thus the Dirac condensate $\left\langle{{\ol{\nu }}_{\tau R}{\ul \nu
}_{\tau L}}\right\rangle$ is not present.  We may suppose instead that the
Majorana condensate $\left\langle{{\ul \nu }_{\tau L}{\ul \nu }_{\tau
L}}\right\rangle$ forms, since it is consistent with the discrete
symmetries of (\ref{h}) and (\ref{i}).  As with other fourth family
condensates this is $SU(3{)}_{H}$ violating and thus does not feed down
mass to ${\nu }_{eL}$ and ${\nu }_{\mu L}$.  This would leave ${\nu }_{\tau
L}$ as the light neutrino associated with the $\tau $.   Unlike all other
members of the third family, operators of the form (\ref{a}) do not
contribute mass to ${\nu }_{\tau L}$.  Instead ${\nu }_{\tau L}$ receives a
naturally small mass from the operator in (\ref{k}).  The neutrinos ${\nu
}_{eL}$ and ${\nu }_{\mu L}$ also receive naturally small mass from
operators of the following form (with possible underlines omitted), also
generated by physics far above 1 TeV.  \be
{\mbox{\boldmath${\ol{Q}}_{L}{U}_{R}{\ol{Q}}_{L}{U}_{R}$}}{\ell }_{L}{\ell
}_{L}~~~,~~~{\mbox{\boldmath${\ol{D}}_{R}{Q}_{L}{\ol{D}}_{R}{Q}_{L}$}}{\ell
}_{L}{\ell }_{L}\ee

We are discussing a situation where isospin is being broken in the fourth
family lepton sector due to the Majorana neutrino mass. But the
contribution to $T$ may not be too large if the fourth family lepton masses
are somewhat smaller than the fourth family quark masses.  This case is
similar to a very recent discussion in the technicolor context~\cite{B}.
That analysis also implies that a negative contribution to $S$ develops if
the fourth family neutrino mass is sufficiently small compared to the
fourth family charged lepton mass.

Note that the condensates $\left\langle{{\nu }_{\tau L}{\ul{\nu }}_{\tau
L}}\right\rangle$ and $\left\langle{N}_{L}{\ul{N}}_{L}\right\rangle$, if
they existed, would have had disastrous consequences for all light neutrino
masses.  But they are fortunately not consistent with the discrete symmetry
in (\ref{h}).  This provides some connection between reasonable neutrino
masses and our picture of massless metafermions.

In summary we have outlined how a sector of massless confined metafermions
can contribute to quark and lepton masses through \su\ invariant
four-metafermion condensates.  We have shown how discrete symmetries can
naturally arise which forbid metafermion masses.  This picture suggests a
peculiar metahadron mass spectrum lying between the third and fourth
families.  There are of course many questions to address before claiming
any kind of realistic description of quark and lepton masses.  Clearly more
description of the physics responsible for the nonrenormalizable
interactions is needed.  It may also be the case that the {\em subdominant}
nonrenormalizable interactions break some of the discrete symmetries we
have discussed, and this may play a role in generating some of the small
masses and mixings in the final quark and lepton mass spectrum.

\vspace{3ex}
\noindent {\Large\bf Acknowledgment}
\vspace{1ex}

This research was supported in part by the Natural Sciences and
Engineering Research Council of Canada.

\end{document}